\documentclass[pra,twocolumn,showpacs]{revtex4}
\usepackage{amsmath,amssymb,mathrsfs}
\usepackage{psfrag}
\usepackage{graphicx}
\usepackage{graphics}
\usepackage{bm}
\usepackage{color}
\usepackage{verbatim,color,ulem}

\newcommand{\beq}{\begin{equation}}
\newcommand{\eeq}{\end{equation}}
\newcommand{\beqa}{\begin{eqnarray}}
\newcommand{\eeqa}{\end{eqnarray}}

\begin{document}

\title{From Narrow to Broad Feshbach Resonances: \\
Condensate Fraction of Cooper Pairs and Preformed Molecules}
\author{Luca Salasnich}
\affiliation{Dipartimento di Fisica e Astronomia ``Galileo Galilei'' 
and CNISM, Universit\`a di Padova, Via Marzolo 8, 35131 Padova, Italy}

\date{\today}

\begin{abstract}
We extend our previous investigations of fermionic condensation 
in broad Feshbach resonances by using the two-channel model 
developed for narrow Feshbach resonances. 
We investigate two crossovers: the BCS-BEC 
crossover by changing the s-wave scattering length and 
the crossover from a narrow to a broad resonance
by changing the atom-molecule coupling. 
At zero temperature we analyze, as a function of both atom-molecule 
coupling and s-wave scattering length, the chemical potential, 
the energy gap, and the condensate fraction of atoms. 
In particular, we predict the contribution of Cooper pairs 
and preformed molecules to the total condensate density 
along the two crossovers. 
\end{abstract}


\maketitle

\section{Introduction}

Nowadays there is a remarkable experimental control over ultracold alkali
atoms interacting through a Feshbach resonance. 
Manipulation of the binding energy through external magnetic 
fields enables experimentalists to evolve clouds 
of two-component fermionic atoms 
from the weakly coupled BCS-like behavior of Cooper pairs to the 
strongly coupled Bose-Einstein 
condensation (BEC) of molecules \cite{exp_MolecularBEC}. The
transition is characterized by a crossover in which, most simply,
the $s$-wave scattering length $a_s$ diverges as it changes sign 
\cite{exp_Crossover,chin}. Recently, a considerable theoretical effort 
\cite{sala-odlro,ortiz,ohashi2,sala-odlro2,sala-odlro3,sala-odlro4,
cinesi,sala-odlro5,sala-odlro6} 
has been expended on studying the condensate fraction of Cooper 
pairs of such a tunable superfluid. The behavior of 
tunable gases can differ according as the Feshbach resonance is 
broad or narrow, the former essentially describing a one-channel system, 
the latter a two-channel system \cite{gur1,gurarie,rivers,ohashi}. 
In two experiments \cite{zwierlein,ueda} with a broad Feshbach resonance  
the condensate fraction of Cooper pairs \cite{yang} 
has been studied in two hyperfine component Fermi vapours of $^6$Li atoms. 
The experimental data of the condensate fraction, which is directly related
to the off-diagonal-long-range order of the two-body density
matrix of fermions \cite{penrose,campbell}, are in quite good agreement
with broad-resonance mean-field theoretical predictions 
\cite{sala-odlro,ortiz} at zero temperature, 
while at finite temperature beyond-mean-field 
corrections are needed \cite{ohashi2}. 

In this paper we extend our previous investigations of fermionic 
condensation in broad Feshbach resonances \cite{sala-odlro,sala-odlro2,
sala-odlro3,sala-odlro4,sala-odlro6}
by analyzing the condensate fraction of Cooper 
pairs and the condensate fraction of preformed molecules 
in the case of a narrow Feshbach resonance. We find that
by increasing the resonant atom-molecule coupling
the condensate fraction of preformed molecules
is strongly reduced in the BEC regime
while the condensate fraction of Cooper-paired atoms grows
reaching a maximum value close to the unitarity limit (infinite
scattering length). Finally, for very large values of the
resonant coupling we recover the broad-resonance regime 
where there are no more preformed molecules
and the BCS-BEC crossover is enterely due to Cooper-paired atoms. 

\section{Effective action and saddle-point approximation}

We adopt the path integral formalism \cite{tempere} 
and consider a mixture of fermionic atoms and 
molecular bosons, in which the fermions, described by the 
complex Grassmann fields $\psi_{\sigma} ({\bf r},\tau )$, 
with spin $\sigma = ( \uparrow , \downarrow )$, can be bound into 
a molecular boson described by the complex scalar field 
$\phi({\bf r},\tau )$ through a Feshbach resonance \cite{gurarie,rivers}. 
The partition function ${\cal Z}$ of the system at temperature $T$ 
can be written as 
\beq 
{\cal Z} = \int {\cal D}\psi_{\sigma} {\cal D}\psi^*_{\sigma}
{\cal D}\phi {\cal D}\phi^* \ 
\exp{\left( -S/\hbar \right) } \; , 
\eeq
where the Euclidean action functional $S$ is given by 
\beq 
S = \int_0^{\hbar\beta} 
d\tau \int_V d^3{\bf r} \ \left( \mathscr{L}_F + 
\mathscr{L}_B + \mathscr{L}_{FB} \right) \;  
\eeq
with 
\beq 
\mathscr{L}_F =
\sum_{\sigma=\uparrow , \downarrow}
\psi^*_{\sigma} \left[ \hbar {\partial\over\partial\tau} 
- \frac{\hbar^2}{2m}\nabla^2 - \mu \right] \psi_{\sigma} 
\eeq
the Lagrangian density of free fermionic atoms, 
\beq
\mathscr{L}_B = 
\phi^* \left[ \hbar {\partial\over \partial \tau}
- \frac{\hbar^2}{4m}\nabla^2 - 2 \mu + 
\epsilon_0 \right] \phi 
\eeq
the Lagrangian density of free bosonic molecules, and 
\beq 
\mathscr{L}_{FB} = 
g \, \left( \phi^* \, \psi_{\downarrow} \, \psi_{\uparrow} 
+ \phi \, \psi^*_{\uparrow} \, \psi^*_{\downarrow} \right) \; . 
\eeq 
the Lagrangian density of fermion-molecule coupling. 
Notice that $\beta=1/(k_B T)$ with $k_B$ the Boltzmann constant 
and $V$ is the volume of the system. 
Moreover, the bound molecular bosons of Feshbach resonance have  
twice the mass of the fermions and a tunable binding energy $\epsilon_0$.

The action functional $S$ is quadratic in the fermionic fields 
$\psi_{\sigma}({\bf r},\tau)$, which can be then integrated out 
exactly obtaining 
\beq 
{\cal Z}= \int {\cal D}\phi {\cal D}\phi^* \
\exp{\left( -S_e/\hbar \right)} \; , 
\eeq
where $S_e$ is the effective action, given by 
\beq
S_e = - \hbar \, Tr[\ln{\left( \mathscr{G}^{-1} \right)}] + 
\int_0^{\hbar\beta} d\tau \int_V d^3{\bf r} \, \mathscr{L}_B 
\label{ss}
\eeq
with $\mathscr{G}^{-1}({\bf r},{\bf r}',\tau , \tau')$ the inverse 
Green-Nambu function, defined as  
\begin{widetext}
\beq
\mathscr{G}^{-1}({\bf r},{\bf r}',\tau , \tau') = 
{1\over \hbar} 
\left( 
\begin{array}{cc}
\hbar {\partial\over\partial\tau} - \frac{\hbar^2}{2m}
\nabla^2 - \mu & g\,\phi({\bf r},\tau ) 
\\
g\, \phi^*({\bf r},\tau ) & \hbar{\partial\over\partial\tau} + 
\frac{\hbar^2}{2m}\nabla^2 + \mu
\end{array} 
\right) \delta({\bf r}-{\bf r}') \, \delta(\tau - \tau') \; .   
\eeq
\end{widetext}
Within the saddle-point approximation \cite{tempere} we consider 
a space and time-independent molecular field, 
i.e. $\phi({\bf r},\tau )=\phi_0$. Without loss of generality 
we suppose that $\phi_0$ is a real number. In this way 
the corresponding saddle-point effective action function $S_{sp}$ 
obtained from Eq. (\ref{ss}) reads 
\beq 
{S_{sp}\over V} = - 2 \int { d^3{\bf k} \over (2\pi)^3 } 
\ln{\left[ 2\cosh{(\beta E_k/2)}\right] } 
+ \hbar \beta ( \epsilon_0 - 2 \mu ) \phi_0^2 \; , 
\eeq
where 
\beq 
E_k = \sqrt{\xi_k^2+g^2\phi_0^2} \;  
\eeq
are the Bogoliubov-like elementary excitations with 
$\xi_k = {\hbar^2k^2/(2m)}-\mu$. 
The saddle-point thermodynamic grand potential $\Omega_{sp}$ is related 
to the saddle-point partition function ${\cal Z}_{sp}$ and to the 
saddle-point effective action $S_{sp}$ by the formula 
\beq 
{\cal Z}_{sp} = \exp{\left( -S_{sp}/\hbar \right) } = 
\exp{\left( -\beta \, \Omega_{sp} \right) } \; . 
\eeq
The value of $\phi_0$ still needs to be determined. 
It is found from extremizing the 
action function $S_{sp}$, or equivalently $\Omega_{sp}$, with respect 
to $\phi_0$ to determine the saddle-point value. Doing so one finds 
\beq 
{\epsilon_0 - 2\mu\over g^2} = \int { d^3{\bf k} \over (2\pi)^3 } 
{\tanh{(\beta E_k/2)}\over 2E_k} \; .   
\label{gap}
\eeq
The integral on the right side of Eq. (\ref{gap}) is formally 
divergent. However, expressing the bare detuning parameter 
$\epsilon_0$ in terms of the effective scattering length $a_s$ 
with the formula \cite{ohashi3}
\beq 
{\epsilon_0 - 2\mu\over g^2} = - {m\over 4\pi\hbar^2 a_s} 
+ \int { d^3{\bf k} \over (2\pi)^3 } 
{m\over {\hbar^2k^2}} 
\eeq
one obtains the regularized equation 
\beq 
- {m\over 4\pi\hbar^2 a_s} = 
 \int { d^3{\bf k} \over (2\pi)^3 } \left( 
{\tanh{(\beta E_k/2)}\over 2E_k} - 
{m\over {\hbar^2k^2}} \right) \; . 
\label{gap-r}
\eeq

The total number density $n$ of atoms in the system 
is instead obtained from the thermodynamic formula 
$n=-{\partial \Omega_{sp}/\partial \mu}$, which gives 
\beq 
n = \int { d^3{\bf k} \over (2\pi)^3 } \left[ 1 - 
(u_k^2 - v_k^2)  \tanh{(\beta E_k/2)} \right] + 2\phi_0^2 \; . 
\label{number}
\eeq
with $u_k^2 = \left( 1 + {\xi_k/E_k} \right)/2$ and 
$v_k^2 = \left( 1 - {\xi_k/E_k} \right)/2$. 
Eq. (\ref{number}) can be written as $n = n_F+n_B$ where $n_B=2\phi_0^2$ 
is the density of atoms in the bosonic molecules 
(with two atoms per molecule). 
We stress that Eqs. (\ref{gap-r}) and (\ref{number}) 
generalize the gap and number equations one finds for 
a broad Feshbach resonance. In fact, setting 
\beq 
\Delta = g \, \phi_0 \; , 
\label{eq16}
\eeq
the broad-resonance regime is easily 
obtained in the limit $g\to +\infty$ with the constraint 
of a finite energy gap $\Delta$ which implies $\phi_0\to 0$. 
Notice that in the literature 
it is common to study models that in addition to
the two-channel Feshbach resonant interaction considered
above, a featureless nonresonant four-Fermi atomic 
interaction is also included \cite{ohashi}. 
Gurarie and Radzihovsky \cite{gurarie} 
have shown that in three dimensions doing 
so does not add any new physics to the pure two-channel 
model considered here. 

\section{Condensation}

The total condensate density $n_0$ of the system 
is the sum of two contributions, 
namely $n_0=n_{F,0}+n_{B,0}$, where 
$n_{F,0}=2|\langle \psi_{\uparrow}({\bf r},\tau )
\psi_{\downarrow}({\bf r},\tau )
\rangle|^2$ is 
the condensate density of Cooper-paired atoms and 
$n_{B,0}=2|\langle \phi({\bf r},\tau ) \rangle|^2$ is 
the condensate density of atoms 
in the preformed bosonic molecules. 
Within the saddle-point approximation we have $n_{B,0}=2\phi_0^2$ 
and taking into account previous results 
\cite{sala-odlro,ortiz,ohashi2,sala-odlro2,sala-odlro3,sala-odlro4,
cinesi,sala-odlro5,sala-odlro6} on the condensate fraction of Cooper-paired 
atoms we finally obtain 
\beq 
n_0 = \int { d^3{\bf k} \over (2\pi)^3 } \left[ u_k^2 v_k^2 
\tanh^2{(\beta E_k/2)} \right] + 2 \phi_0^2 \; . 
\label{condensate}
\eeq

\begin{figure}[t]
\begin{center}
{\includegraphics[width=8.cm,clip]{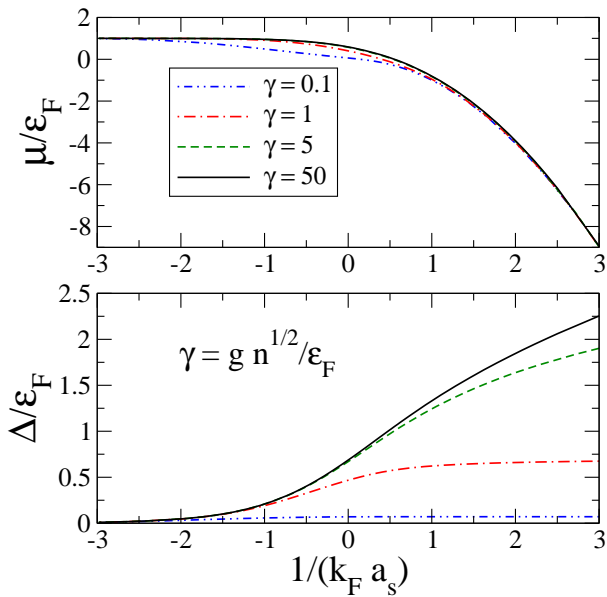}}
\end{center}
\caption{(Color online). Upper panel: Scaled chemical
potential $\mu/\epsilon_F$ as a function of the scaled inverse scattering
length $1/(k_Fa_s)$. Lower panel: Scaled energy gap $\mu$ as a function
of the scaled inverse scattering length $1/(k_Fa_s)$.
The curves correspond to different values of the scaled 
atom-molecule coupling $\gamma=gn^{1/2}/\epsilon_F$.}
\label{fig1}
\end{figure}

As discussed in \cite{gurarie}, at zero temperature 
the saddle-point approximation gives reliable results  
while at finite temperature it is necessary 
to include fluctuations about the saddle point $\phi_0$: 
fluctuations are particularly important 
close to the critical temperature $T_c$ of 
the super-to-normal phase transition. For this reason we analyze 
Cooper pairing and Bose-Einstein condensation in narrow Feshbach 
resonances by using Eqs. (\ref{gap-r}), (\ref{number}) 
and (\ref{condensate}) only at zero temperature, where they reduce 
to the following ones: 
\beqa
y &=& - {2\over \pi} I_1\Big({{\tilde \mu}\over {\tilde \Delta}}\Big)  \; , 
\label{pp1}
\\
1 &=& {3\over 2} {\tilde \Delta}^{3/2}
I_2\Big({{\tilde \mu}\over {\tilde \Delta}}\Big) 
+ 2 {{\tilde \Delta}^2\over \gamma^2} \; , 
\label{pp2}
\\
{n_0\over n} &=& {3\pi\over 2^{7/2}} {\tilde \Delta}^{3/2} 
\sqrt{{{\tilde \mu}\over {\tilde \Delta}}+\sqrt{1+{{\tilde \mu}
\over {\tilde \Delta}}}} 
+ 2 {{\tilde \Delta}^2\over \gamma^2} \; ,
\label{pp3}
\eeqa
where $\tilde{\mu}=\mu/\epsilon_F$ is the adimensional chemical 
potential with $\epsilon_F=\hbar^2k_F^2/(2m)$ the Fermi energy and 
$k_F=(3\pi^2n)^{1/3}$ is the Fermi wavenumber, 
$\tilde{\Delta}$ is the adimensional energy gap, 
$y=1/(k_Fa_s)$ is the adimensional inverse scatering length, 
$\gamma = g n^{1/2}/\epsilon_F$ is the adimensional resonant atom-molecule 
coupling, $I_1(x)$ and $I_2(x)$ are the two monotonic functions 
\beqa 
I_1(x) &=& \int_0^{+\infty} z^2 
\left( {1\over \sqrt{(z^2-x)^2+1}} - {1\over z^2} \right) dz \; , 
\\
I_2(x) &=& \int_0^{+\infty} z^2 \left( 1 - {z^2-x\over \sqrt{(z^2-x)^2+1}}
\right) dz \; , 
\eeqa
which can be expressed in terms of elliptic integrals. 

We have solved Eqs. (\ref{pp1}) and (\ref{pp2}) numerically 
obtaining the scaled (adimensional) chemical 
potential $\mu/\epsilon_F$ and the scaled energy gap $\Delta/\epsilon_F$ 
as a function of $y=1/(k_Fa_s)$ for different values 
of $\gamma=g n^{1/2}/\epsilon_F$. The results are shown 
in Fig. \ref{fig1}. Remarkably the chemical potential $\mu$ does 
not depend very much on the resonant atom-molecule coupling $\gamma$ 
apart close to the unitarity region ($-1<y<1$). Instead the behavior 
the energy gap $\Delta$ depends strongly on $\gamma$ in the BEC region 
($y>1$). In particular, in the limit $y\to +\infty$ the energy gap $\Delta$ 
goes to a value which is obtained as the solution of Eq. (\ref{pp2}) 
with $I_2(-\infty)=0$, namely $\Delta/\epsilon_F =\gamma/\sqrt{2}$. 
Taking into account Eq. (\ref{eq16}), this result means that in the limit 
$y\to+\infty$ one always has $n_{B,0}=n$, for any finite value of $\gamma$. 

\begin{figure}[t]
\begin{center}
{\includegraphics[width=8.cm,clip]{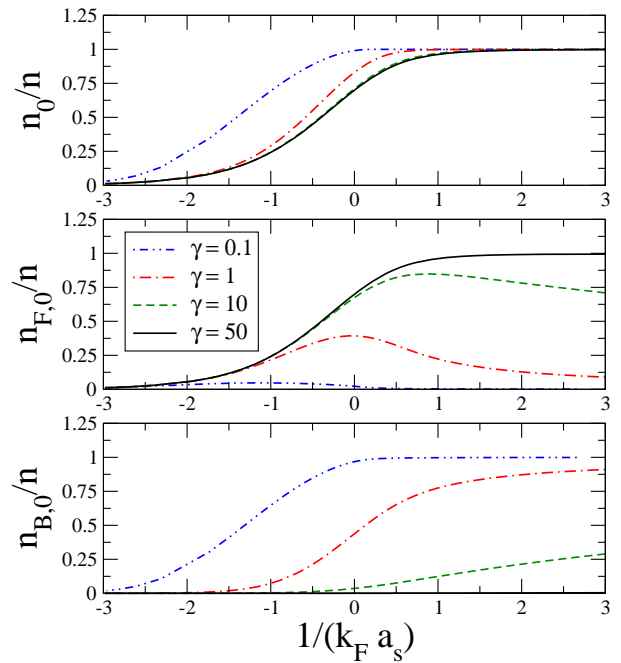}}
\end{center}
\caption{(Color online). Upper panel: Total condensate 
fraction $n_0/n$ as a function of the scaled inverse scattering 
length $1/(k_Fa_s)$. Middle panel: Condensate 
fraction $n_{F,0}/n$ of Cooper paired atoms 
as a function of the scaled inverse scattering 
length $1/(k_Fa_s)$. Lower panel: Condensate 
fraction $n_{B,0}/n$ of atoms in preformed molecules 
as a function of the scaled inverse scattering length $1/(k_Fa_s)$. 
The curves correspond to different values of the scaled 
atom-molecule coupling $\gamma=gn^{1/2}/\epsilon_F$.}
\label{fig2}
\end{figure}

In the upper panel of Fig. \ref{fig2} we plot the total condensate 
fraction $n_0/n$, obtained from Eq. (\ref{pp3}), 
as a function of the scaled inverse scattering length $1/(k_Fa_s)$ 
for four values of $\gamma$. For a very small atom-molecule coupling, 
i.e. $\gamma=0.1$ (dot-dot-dashed curves), the total condensate 
fraction becomes equal to $1$ at unitarity $y=0$. Indeed in this 
regime of weak-coupling the fraction $n_0/n$ is mainly due 
to preformed molecules, 
as clearly shown in the middle panel, where we plot 
the condensate fraction $n_{F,0}/n$ of Cooper paired atoms, 
and in the lower panel, where we plot the condensate 
fraction $n_{B,0}/n$ of atoms in preformed molecules. 
By increasing the resonant coupling $\gamma$, see for instance 
dot-dashed ($\gamma=1$) and dashed ($\gamma=10$) curves, 
the value of the total condensate fraction at unitarity ($y=0$)
decreases. Moreover, by increasing $\gamma$ 
in the BEC region ($y>1$) it appears a finite 
condensate fraction of Cooper-paired (and bound) atoms while 
the fraction of condensed preformed molecules is strongly reduced. 
Finally, for very large values of $\gamma$, e.g. $\gamma=50$ 
(solid curves) of Fig. \ref{fig2}, in practice (i.e. in a 
large range of $y$) one recovers the broad-resonance 
limit where there are not preformed molecules and the BCS-BEC crossover 
is enterely due to Cooper-paired atoms. 

Within the zero-temperature 
saddle-point approximation we have used, the condensate density 
$n_{F,0}$ of fermions is smaller than the total density $n_F$ 
of fermions, while the condensate density 
$n_{B,0}$ of preformed molecules coincides with the total density 
$n_B$ of preformed molecules. In general, as shown by 
Werner, Tarruell, and Castin \cite{castin}, for the two-channel 
model the density $n_B$ (but not $n_{B,0}$ nor $n_{F,0}$) can be directly 
related to the derivative of the energy of the gas with respect 
to the inverse scattering length. The zero-temperature 
condensate depletion $(n_{B}-n_{B,0})/n_B$ of bosonic molecules 
can be instead obtained going beyond the saddle-point approximation 
by including quadratic fluctuations of $\phi({\bf r},\tau)$ 
around $\phi_0$ in the action functional. 

\section{Conclusions} 

In this paper we have used the two-channel model of atomic 
fermions coupled to preformed bosonic molecules 
to study two crossovers: the BCS-BEC 
crossover driven by the s-wave scattering length and
the crossover from a narrow to a broad resonance 
controlled by the atom-molecule coupling. 
Within the saddle-point approximation of the 
path integral formalism we have derived an 
extended gap equation which depends on both 
the s-wave scattering length and the atom-molecule coupling. 
Only for a very large atom-molecule coupling the extended 
gap equation reduces to the one of broad Feshbach resonances. 
At zero temperature, we have calculated the chemical potential 
and the energy gap of the system. Remarkably, in the deep BEC region of the 
BCS-BEC crossover the energy gap $\Delta$ goes to the asymptotic 
value $g\, n^{1/2}/\sqrt{2}$, where $g$ 
the atom-molecule coupling and $n$ is the total atomic 
density. We have then analyzed the condensate density and 
the condensate fraction of the system. We have found that
by increasing the resonant atom-molecule coupling 
the condensate fraction of preformed molecules
is strongly reduced in the BEC regime with a corresponding 
growth in the condensate fraction of Cooper-paired atoms. 
As expected, for very large values of the 
resonant coupling one recovers the broad-resonance regime 
characterized by the absence of preformed molecules.

\end{document}